\documentstyle[psfig,epsf,conf-X]{article}
\begin{document} 
\small
\enlargethispage{1truein}
\vskip-1.0truein
\noindent To appear in {\it Large Scale Structure in the X-ray Universe},
editted by M. Plionis \& I. Georgantopoulos (Paris: Editions Frontieres),
in press.

\vskip0.7truein
\heading{%
%Begin Heading
%
Thermal and Nonthermal Effects of Merger Shocks in Clusters of Galaxies
% End Heading
}
\par\medskip\noindent
\author{%
%Begin Author names
Craig L. Sarazin$^1$
%End Author names
}
\address{%
%First address
Department of Astronomy, University of Virginia,
P.O. Box 3818, \\
Charlottesville, VA 22901, USA; cls7i@virginia.edu
}

\begin{abstract}
Cluster mergers drive shocks into the intracluster medium which
heat and compress the thermal gas.
X-ray observations of shocks can be used to determine the geometry and
kinematics of the merger.
Merger shocks should also accelerate relativistic particles.
Electrons with $\gamma \sim 300$ ($E \sim 150$ MeV) are expected to
be particularly common.
Relativistic particles produce observed EUV emission, hard X-ray tails,
and diffuse radio emission.
The predicted gamma-ray fluxes of clusters should make them easily
observable with GLAST.
\end{abstract}
\section{Introduction}
Major cluster mergers are the most energetic events in the Universe
since the Big Bang.
% Cluster mergers are the mechanism by which clusters are assembled.
In these mergers, the subclusters collide at velocities of
$\sim$2000 km/s, and shocks are driven into the intracluster medium.
In major mergers, these hydrodynamical shocks dissipate energies of
$\sim 3 \times 10^{63}$ ergs; such shocks are the major heating
source for the X-ray emitting intracluster medium.
% The shock velocities in merger shocks are similar to those in
% supernova remnants in our Galaxy, and we expect them to produce similar
% effects.
Mergers shocks should heat and compress the X-ray emitting intracluster
gas, and increase its entropy.
We also expect that particle acceleration by these shocks will produce
nonthermal electrons and ions, and these can produce synchrotron
radio, inverse Compton (IC) EUV and hard X-ray, and gamma-ray emission.

\begin{figure}
\includegraphics{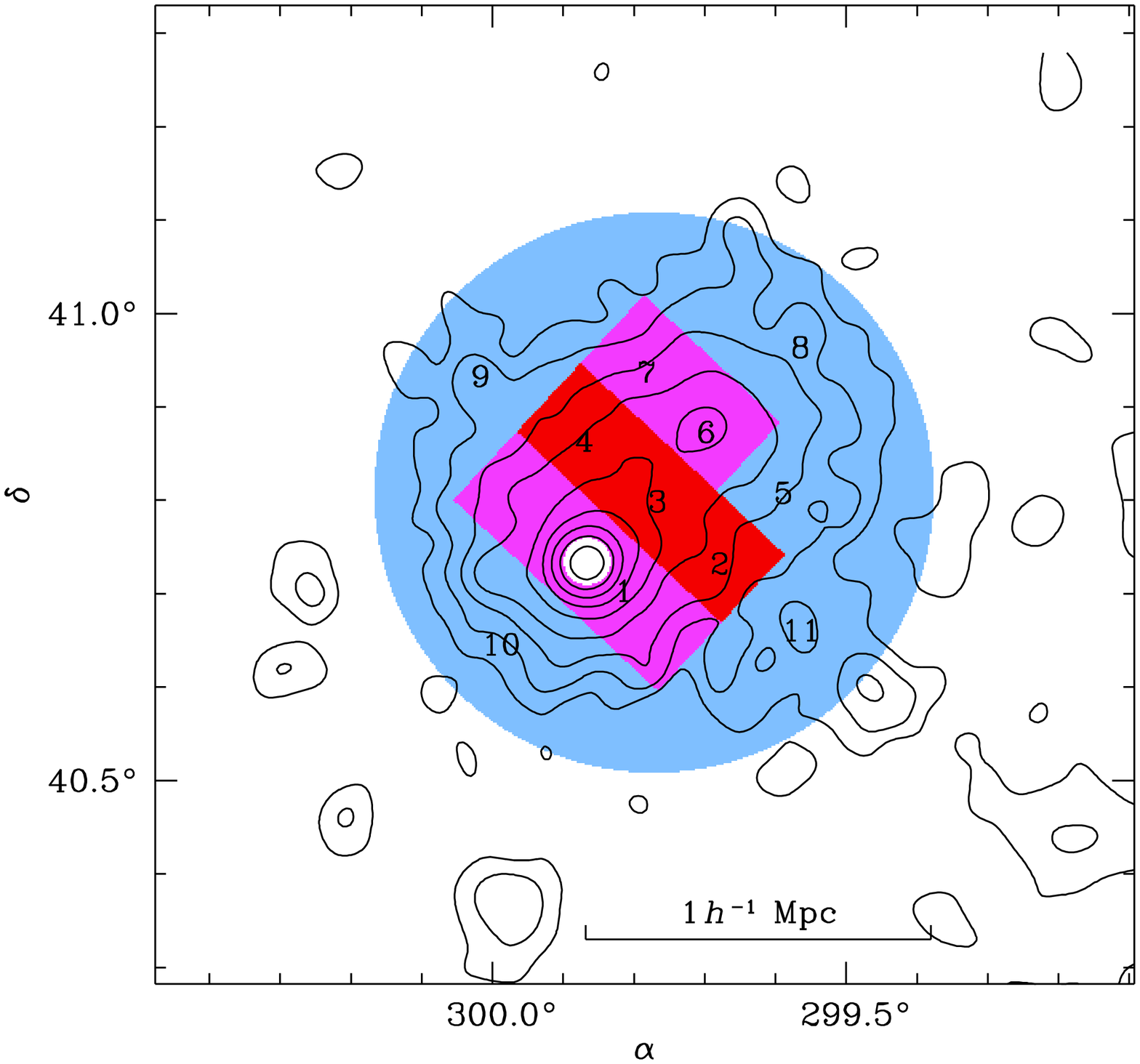}
\includegraphics{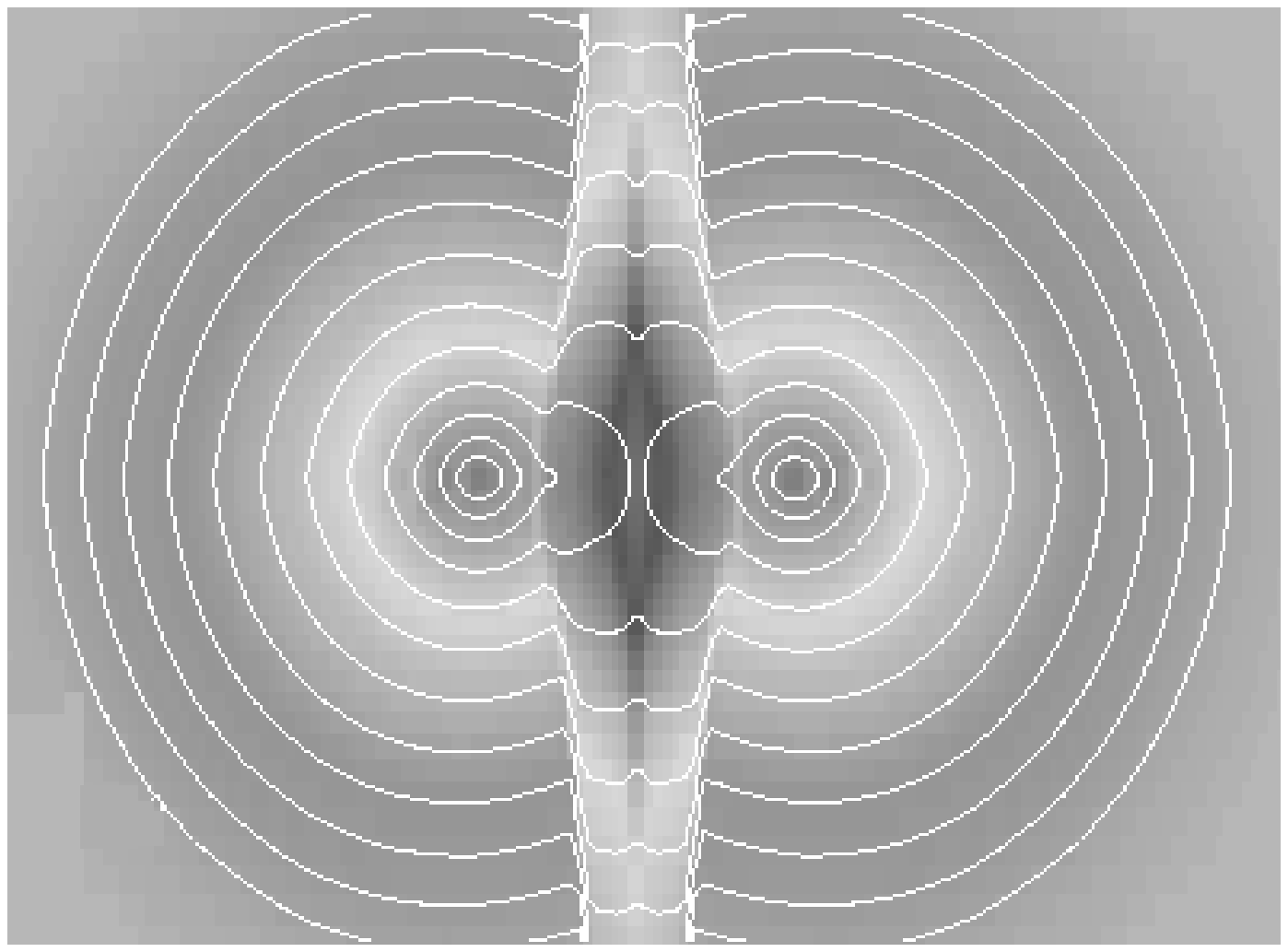}
\centerline{\null}
\centerline{\null}
\noindent\hskip1.6truein (a) \hskip2.40truein (b) \hfill
\vskip1.6in
%\vskip1.9in
\caption[]{(a) ROSAT PSPC contours of the Cygnus A cluster.
The radio galaxy is located at the center of the cooling flow peak in
the southeast subcluster.
A second subcluster is centered on the X-ray peak 11$^\prime$ to the
northwest.
The greyscale shows the ASCA temperatures, and indicates that a merger
shock is located between the two subcluster centers\cite{MSV}.
(b)
A hydro/N-body simulation of the Cygnus A cluster merger\cite{RS}.
\label{fig:cyga}
}
\end{figure}

\section{Thermal effects of merger shocks}
Merger shocks heat and compress the intracluster gas, and these effects
can be used to determine the geometry and kinematics of the merger.
ASCA X-ray temperature maps and ROSAT images have been used in an
initial effort to apply this technique\cite{MSV}.
The cluster containing the radio source Cygnus A is a particularly simple
case (Fig.~\ref{fig:cyga}).
This appears to be a fairly symmetric merger with a low impact parameter.
The merger is at an early phase, with the merger shocks being located
between the two subcluster centers.
A hydro/N-body simulation of the merger\cite{RS} is shown at the right in
Fig.~\ref{fig:cyga} (not to the same scale).
Presumably, the fact that the merger shocks have not yet passed through
the subcluster centers is the reason why the merger hasn't disrupted the
Cygnus A radio source or the surrounding cooling flow.
The simple geometry of this merger makes it easy to apply the shock
jump conditions to determine the merger velocity.
From the Rankine--Hugoniot jump conditions, the velocity change across
the merger shock is 
\begin{equation}
\Delta u_{sh} = \left[\frac{k T_0}{\mu m_p}\left(r-1\right)\left(
\frac{T_1}{T_0}-\frac{1}{r}\right)
\right]^{1/2} \, , \label{eq:shock1}
\end{equation}
where $T_0$ and $T_1$ are the pre- and post-shock temperature,
% $\mu$ is the mean mass per particle,
and the shock compression $r$ is
given by
\begin{equation}
\frac{1}{r} =
\left[\frac{1}{4}\left(\frac{\gamma+1}{\gamma-1}\right)^2
\left(\frac{T_1}{T_0}-1\right)^2 +\frac{T_1}{T_0}\right]^{1/2}
-\frac{1}{2}\frac{\gamma+1}{\gamma-1}\left(\frac{T_1}{T_0}-1\right).
\label{eq:shock2}
\end{equation}
For a symmetric merger, the merger velocity of the two subclusters is
just $\Delta u_{cl} = 2 \Delta u_{sh}$.
When the ASCA temperatures are used, this gives
$\Delta u_{cl} \approx 2200$ km/s.
The radial velocity distribution of the galaxies in this cluster
is bimodal\cite{OLMH} and consistent with a merger velocity of 2400 km/s.

Interestingly, the collision velocity found above is close to the free-fall
% velocity of $\sim2200$ km/s that the two subclusters should have achieved
% had they fallen from a large distance to their observed separation.
velocity of $\sim$2200 km/s for the two subclusters 
if they had fallen from a large distance to their observed separation.
This consistency suggests that the shock energy is effectively thermalized,
and that a major fraction does not go into turbulence, magnetic fields,
or cosmic rays. 

% Cluster merger are expected to produce collisionless shocks, as occurs
% in supernova remnants.
% As such, nonequilibrium effects are expected, including nonequipartition
% of electrons and ions and nonequilibrium ionization\cite{MSV},\cite{Tak}.
% In fact, observations of a central shock in the spectacular merging
% cluster Abell 3667 show an apparent lag of $\sim$200 kpc between the
% shock position and % the peak in the electron temperature\cite{MSV}.
% This is consistent with weak electron heating at the shock, and the
% timescale for Coulomb heating.

\begin{figure}
\includegraphics{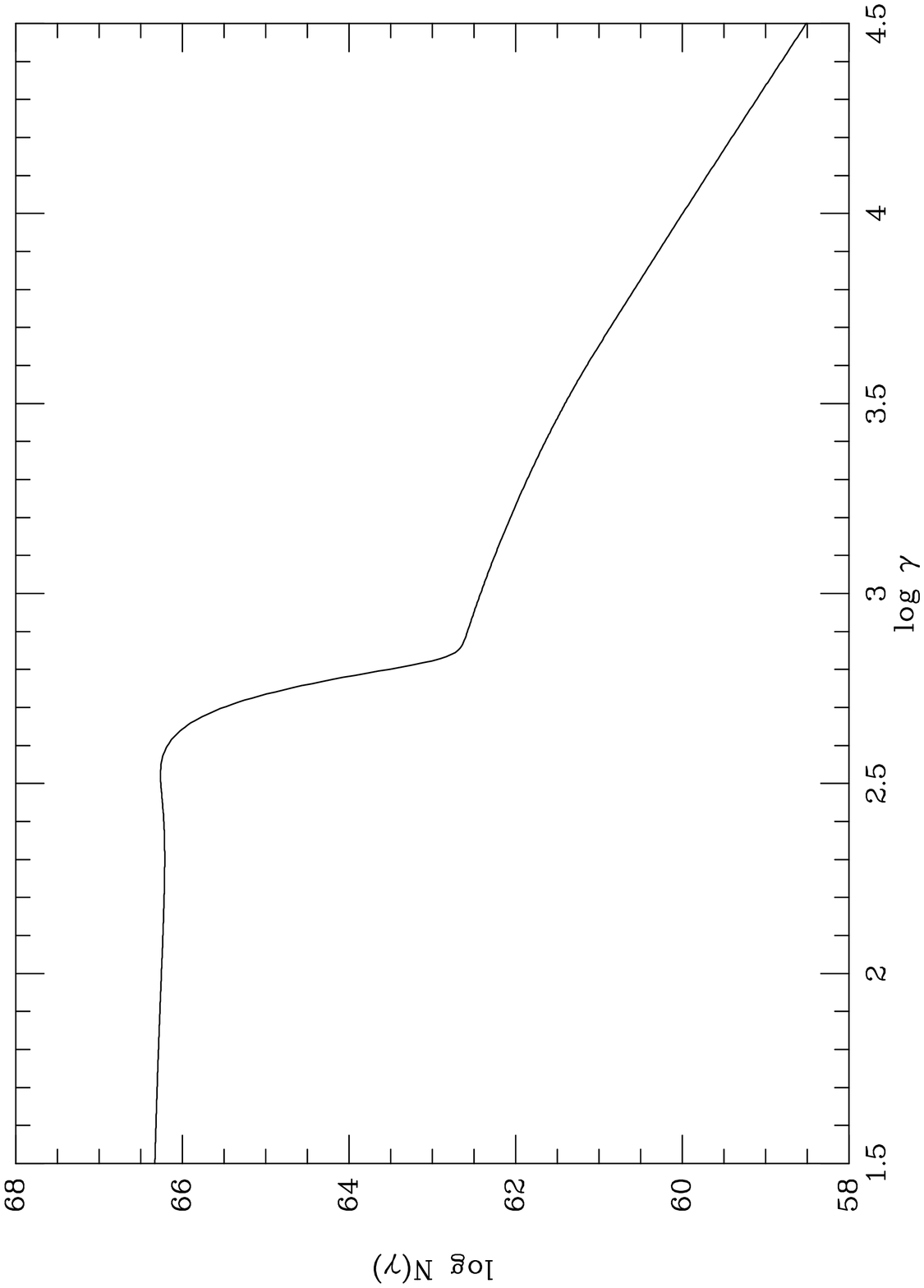}
\includegraphics{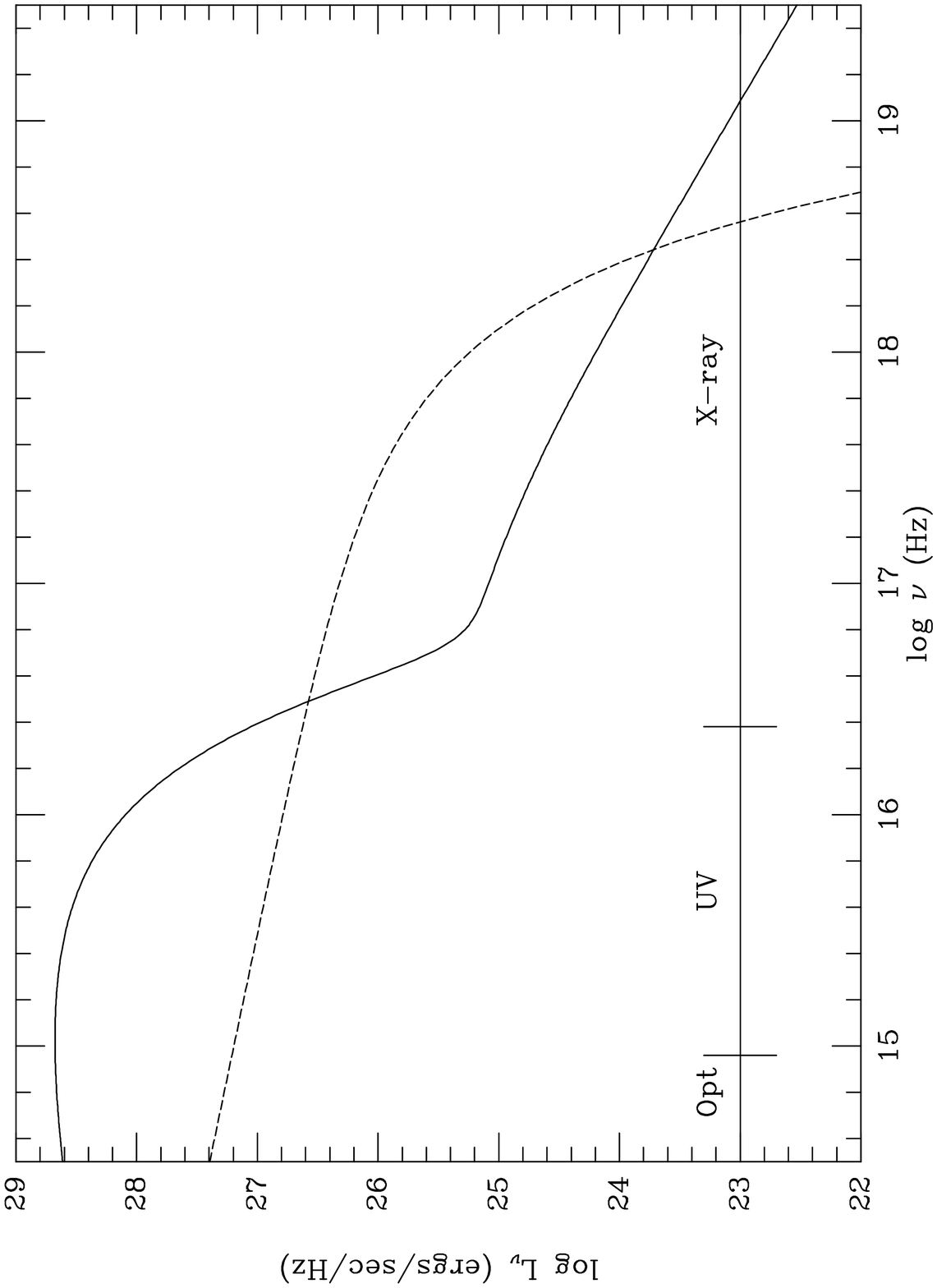}
\centerline{\null}
\centerline{\null}
\noindent\hskip1.9truein (a) \hskip2.10truein (b) \hfill
\vskip1.1in
\caption[]{(a) A typical model for the relativistic electron population
in a cluster of galaxies.
The lower energy electrons are due to all of the mergers in the cluster
history, while the high energy electrons are due to a small current
merger.
(b)
The IC spectrum from the same model (solid curve).
The dashed curve is a 7 keV thermal bremsstrahlung spectrum.
\label{fig:espect}
}
\end{figure}

\section{Nonthermal effects of merger shocks}
Radio observations of supernova remnants indicate that at least a few
percent of the shock energy goes into the acceleration of relativistic
electrons in shocks with $v \gsim 10^3$ km/s.
Even more energy may go into relativistic ions.
Given that all of the thermal energy of the intracluster gas in clusters
is due to shocks with such velocities, it seems likely that relativistic
electrons with a total energy of $\gsim 10^{62}$ ergs are produced in
clusters, with perhaps even higher energies in ions.
Clusters are also very good storage locations for cosmic rays.
Under reasonable assumptions for the diffusion coefficient, particles
with energies $\lsim$$10^6$ GeV have diffusion times which
are longer than the Hubble time\cite{CB}.
Although high energy electrons lose energy rapidly due to IC and
synchrotron emission, electrons with Lorentz factors of $\gamma \sim 300$
have long lifetimes\cite{S1},\cite{SL}.
Thus, clusters of galaxies can retain low energy electrons
($\gamma \sim 300$) and nearly all cosmic ray ions for a significant
fraction of a Hubble time.

Recently, I have calculated models for the relativistic electrons in
% clusters, assuming they are primary electrons accelerated in merger
% shocks\cite{S1},\cite{S2}.
% clusters, assuming they are primary\cite{S1},\cite{S2}.
clusters, assuming they are primary\cite{S1}.
Fig.~\ref{fig:espect}(a) shows the electron spectrum in a cluster
with a typical history.
Most of the electron energy is in electrons with $\gamma \sim 300$, which
have the longest lifetimes.
These electrons are produced by mergers over the entire history
of the cluster.
This cluster also has a small ongoing merger which produces
the high energy tail on the electron distribution.

Most of the emission from these electrons is due to IC, and the
resulting spectrum is shown in Fig.~\ref{fig:espect}(b).
For comparison, thermal bremsstrahlung with a typical rich cluster
temperature and luminosity is shown as a dashed curve.
Fig.~\ref{fig:espect}(b) shows that clusters should be strong
sources of extreme ultraviolet (EUV) radiation.
Since this emission is due to electrons with $\gamma \sim 300$ which
have very long lifetimes, EUV radiation should be a common feature
of clusters\cite{SL}.
In fact, the EUVE satellite appears to have detected all of the clusters
it observed\cite{K},\cite{L1},\cite{L2},\cite{MLL}.
% with long observations and low Galactic
% columns\cite{K},\cite{L1},\cite{L2},\cite{MLL}.
% columns\cite{K},\cite{L1},\cite{MLL}.

\setlength{\rightskip}{2.3truein}
In clusters with an ongoing merger, the higher energy electrons will
produce a hard X-ray tail;
the same electrons will produce diffuse radio synchrotron emission.
Such radio halos are seen in a number of clusters, all of which show
evidence for ongoing mergers\cite{G}.
BeppoSAX observations show hard X-ray excesses in the Coma\cite{FF}
and
Abell 2199\cite{K} clusters.
Coma has a radio halo and is undergoing at least one merger;
on the other hand,
Abell 2199 has no radio halo\cite{KS} nor any evidence for a
merger.
It is possible that the hard tail in Abell 2199 has some other
explanation;
for example, it might be due to nonthermal bremsstrahlung\cite{SK}.

\setlength{\rightskip}{0.0truein}

\hskip-0.2in
\vskip-2.2in
\begin{minipage}{2.2truein}
\end{minipage}
\hskip 2.1truein
\begin{minipage}{2.2truein}
\vskip1.2truein
\includegraphics{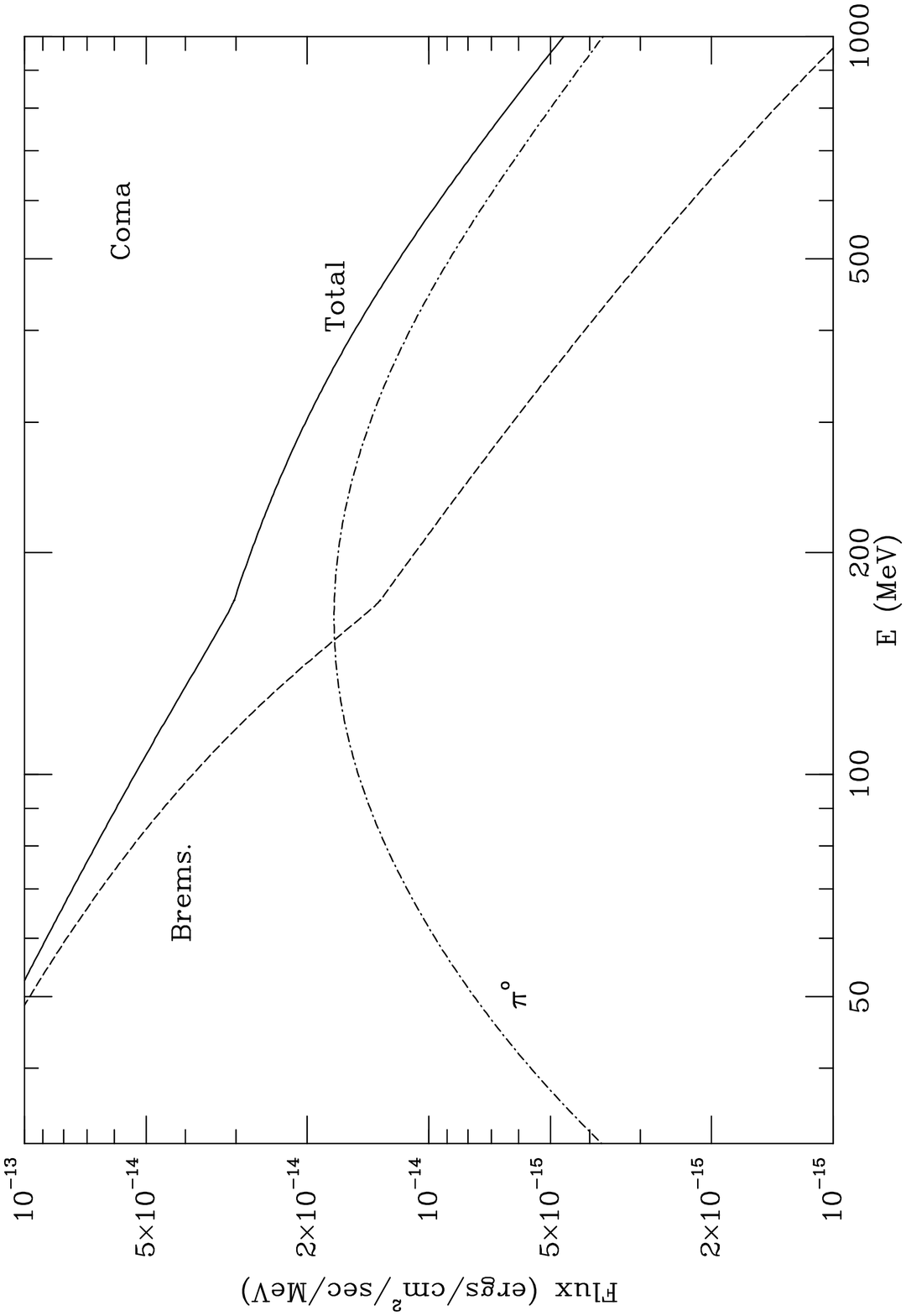}
\noindent Figure 3:
The predicted gamma-ray spectrum for the Coma cluster, including
% electron bremsstrahlung and $\pi^o$ decay from ions\cite{S2}.
electron bremsstrahlung and $\pi^o$ decay from ions.
\end{minipage}

\vskip0.03truein
\vskip\baselineskip
\vskip\baselineskip

Relativistic electrons and ions in clusters are also expected to produce
strong gamma-ray emission.
The region near 100 MeV is particularly
interesting, as this region includes bremsstrahlung from the most common
electrons with $\gamma \sim 300$ and gamma-rays from ions produced by
$\pi^o$ decay.
The predicted fluxes are such that nearby clusters should be easily
detectable with GLAST.

\acknowledgements{This work was done in collaboration with
Josh Kempner, Maxim Markevitch, Paul Ricker, and Alexey Vikhlinin.
It was supported by NASA grant NAG5-8390.}

\vfill
\end{document}